\documentclass[12pt]{article}
\usepackage{epsfig}
\usepackage{pst-plot,url,epsf}
\newcommand{\beq}{\begin{equation}}
\newcommand{\eeq}{\end{equation}}
\newcommand{\bea}{\begin{eqnarray}}
\newcommand{\eea}{\end{eqnarray}}
\newcommand{\beas}{\begin{eqnarray*}}
\newcommand{\eeas}{\end{eqnarray*}}
\newcommand{\nn}{\nonumber}
\newcommand{\epm}{e^+e^-}

\newcommand{\ra}{\rightarrow}

\newcommand{\AmS}{{\protect\the\textfont2
  A\kern-.1667em\lower.5ex\hbox{M}\kern-.125emS}}

\begin{document}
\begin{center}
{\LARGE\bf 
New developments in {\tt carlomat}\footnote{Presented at 
the XXXV International Conference of Theoretical Physics, `` Matter to the Deepest'', Ustro\'n, Poland, September 12--18, 2011.}
}\\
\vspace*{2cm}
K. Ko\l odziej\footnote{E-mail: karol.kolodziej@us.edu.pl}\\[1cm]
{\small\it
Institute of Physics, University of Silesia\\ 
ul. Uniwersytecka 4, PL-40007 Katowice, Poland}\\
\vspace*{2.5cm}
{\bf Abstract}\\
\end{center}
New developments in {\tt carlomat}, a program for automatic 
computation of the lowest order cross sections, are presented. They include 
improvements of the phase integration routines and implementation of 
extensions of the standard model, such as scalar electrodynamics or the
anomalous $Wtb$ coupling including operators of dimension up to five.

\section{Introduction}

{\tt carlomat} 
is a program for automatic computation of the lowest order
cross sections, dedicated in particular for the description of 
multiparticle processes of the form
\bea
\label{process}
p_1 +p_2 \;\ra\; p_3 + ... + p_n
\eea
with the maximum number of external particles $n=12$. In (\ref{process}), 
particles have been symbolized by their four momenta in the centre of mass
system (c.m.s.).
The program is written in Fortran 90/95. It generates the matrix element 
for a user specified process 
together with different phase space parametrizations which are used for the 
multichannel Monte Carlo integration of the lowest order cross sections and
event generation.

Version 1.0 of {\tt carlomat} was released 2 years ago \cite{carlomat}.
Since then the program has been successfully used for calculating
cross sections of many different processes,
as, e.g., all standard model (SM) processes of the form
\bea
\label{ee8f}
e^+ e^- \ra b f_1 \bar{f'_1} \bar{b} f_2 \bar{f'_2} b\bar b,
\eea
where
$f_1, f'_2 =\nu_{e}, \nu_{\mu}, \nu_{\tau}, u, c$ and
$f'_1, f_2 = e^-, \mu^-, \tau^-, d, s$ \cite{KS}
that are relevant for the associated production an decay of a top 
quark pair and a light Higgs boson at the $\epm$ linear collider
\cite{ILC}, \cite{CLIC}.
There are 240\,966 Feynman diagrams for the hadronic channel of (\ref{ee8f})
\bea
\label{ee8fhad}
\epm \;\ra\; b c\bar s \bar b s \bar c b \bar b,
\eea
in the unitary gauge, assuming vanishing masses of light particles
$m_e=m_s=0$, and neglecting the Cabibbo-Kobayashi-Maskawa (CKM) mixing.
Because of so many Feynman diagrams, the matrix element $M$ of
process (\ref{ee8fhad}),
which is calculated in the helicity base, is rather complicated.
However, if the Monte Carlo (MC) summing over helicities is applied,
calculating $\overline{\left|M\right|^2}$
is not a problem in practice. The main issue is to calculate the integral over
$8\times 3 -4 = 20$ 
dimensional phase space of process (\ref{ee8fhad}).

\section{Phase space integration in {\tt carlomat}}

The phase space integration in {\tt carlomat} is performed according to the 
following algorithm. First , final state particles 
$\left\{p_3,p_4,\ldots, p_n\right\}$ of process (\ref{process})
are divided into two subsets of four momenta $q_{i_1}$ and $q_{i_2}$ each,
that are defined in the relative centre of mass system (r.c.m.s.),
$\vec{q}_{i_1}+\vec{q}_{i_1}=\vec{0}$.
This is done in a way that depends on the topology of a diagram.
Then  the identity 
\bea
\int{\rm d}s_i\int\frac{{\rm d}^3 q_i}{2E_i} \; 
\delta^{(4)}\left(q_i-q_{i_1}-q_{i_2}\right)=1,
\qquad E_i^2=s_i+{\vec q}^2_i
\eea
is inserted in the standard parametrization of the Lorentz invariant phase 
space element 
\bea
\label{dlips}
{\rm d}^{3n_f-4} Lips  = (2\pi)^{4} \delta^{(4)}\Big(p_1 +p_2 
- \sum_{i=3}^n p_i\Big) \prod_{i=3}^n\frac{{\rm d} p_i^3}{(2\pi)^32E_i},
\eea
where $n_f=n-2$ is the number of final state particles of process
(\ref{process}).
The insertion is repeated consecutively until parametrization (\ref{dlips}) 
is brought into the following form
\bea
\label{psparam}
 {\rm d}^{3n_f-4} Lips  = (2\pi)^{4-3n_f}  
          {\rm d} l_{0} {\rm d} l_{1}...{\rm d} l_{n-4}
          {\rm d} s_{1} {\rm d} s_{2}...{\rm d} s_{n-4},
\eea
where invariants $s_i$ are given by
\bea
\label{invs}
s_i=\left\{ \begin{array}{l}
\left(q_{i_1}+q_{i_2}\right)^2=\left(E_{i_1}+E_{i_2}\right)^2, \quad {\rm for}
\;\; i=1,...,n-4 \\
\left(p_1+p_2\right)^2=s, \quad {\rm for}\;\; i=0 
\end{array} \right.
\eea
and ${\rm d} l_i$ are the two particle phase space elements 
\bea
\label{2pps}
{\rm d} l_i=\frac{\lambda^{\frac{1}{2}}
\left(s_i,q_{i_1}^2,q_{i_2}^2\right)}{2\sqrt{s_i}}
 {\rm d} \Omega_i.
\eea
In Eq. (\ref{2pps}), 
$\lambda(x,y,z)=x^2+y^2+z^2-2xy-2xz-2yz$ is the kinematical function and
$\Omega_i$ is the solid angle
of momentum ${\vec q}_{i_1}$ in the r.c.m.s.

In {\tt carlomat} v. 1.0, a different phase space parametrization 
(\ref{psparam})
is generated for each of $N$ Feynman diagrams of process (\ref{process})
\bea
\label{fi}
f_i(x)={\rm d}^{3n_f-4} Lips_i\left(x\right) \qquad i=1,\ldots,N,
\eea
where $x=\left(x_1,...,x_{3n_f-4}\right)$ are uniformly distributed random 
arguments and the normalization condition
\bea
\int\limits_0^{\; 1}{\rm d} x^{3n_f-4}f_i(x)={\rm vol}(Lips)
\eea
is satisfied for each parametrization.
Invariants $s_i$ of (\ref{invs}) are randomly generated within their
physical limits 
which are deduced from 
the topology of the Feynman diagram.
They are generated either according to the uniform distribution
or, if necessary, mappings of
the Breit-Wigner shape of the propagators of unstable particles and
$\sim 1/s$ behaviour of the propagators of massless particles
are performed.
An option is included in the program that allows to
turn on the mapping if the unstable particle decays into 2, 3, 4, ...
on-shell particles. 
Different phase space parametrizations obtained in this
way can be used for testing purposes.

All the parametrizations $f_i(x)$ of Eq.~(\ref{fi}) are then automatically
combined into a single multichannel probability distribution 
\bea
f(x)=\sum_{i=1}^N a_i f_i(x),
\eea
with non negative weights $a_i$, $i=1,...,N$, 
satisfying the condition
\bea
\sum_{i=1}^N a_i = 1
\qquad \Leftrightarrow\qquad
\int\limits_0^{\; 1}{\rm d} x^{3n_f-4}f(x)={\rm vol}(Lips).
\eea
The actual MC integration is done with the random numbers generated
according to probability distribution $f(x)$. A large number of
the Feynman diagrams, which is typical for multiparticle processes, results in
the equal number of kinematical routines that are generated. The kinematical
routines that contribute the most to the integral can be selected
with the iterative approach described below.

Integration in {\tt carlomat} can be performed iteratively.
First, the MC integral is calculated $N$ times with a rather small number 
of calls to the integrand,
each time with a different phase space parametrization $f_i(x)$, all the
parametrizations being assigned weight $a_i=1/N$.
The result $\sigma_i$ obtained with $i$-th parametrization is used to
calculate a new weight according to the following formula
\bea
          a_i=\sigma_i/\sum_{j=1}^N \sigma_j
\eea
that is the probability of choosing $i$-th parametrization in the first 
iteration.
In this way channels with small weights $a_i$ are not 
chosen and will have zero weights in the next iteration.
After the first iteration has been completed,
the new weights for the second iteration are determined analogously and 
so on.
After several iterations only the most important kinematical channels
survive. However, the large number of kinematical channels for multiparticle
processes in the beginning 
implies a very long compilation time.

This has been improved in the current version of {\tt carlomat} by introducing
the following changes:
\begin{itemize}
\item Commands for calculating phase space boundaries and boosts of the 
particle four momenta from the r.c.m.s, where they randomly generated, 
to the c.m.s. are generated only once for all parametrizations
corresponding to diagrams of the same topology.
\item Kinematical routines corresponding to the diagrams of the same 
topology that 
contain the same mappings are discarded at the stage of code generation.
\end{itemize}
In this way, the generated code is substantially shortened and
reduction of a compilation time, 
typically by a factor $2-5$ for multiparticle processes, is achieved.

\section{Extensions of SM}

Extensions of SM that have been implemented in the current version of 
{\tt carlomat} include scalar electrodynamics
and an anomalous $Wtb$ coupling. The details on the latter are described below.

The most general effective Lagrangian of the $Wtb$ interaction 
containing operators of dimension four and five that has been implemented
in the program has the following form
\cite{kane}
\bea
\label{wtblagr}
L_{Wtb}&=&\frac{g}{\sqrt{2}}\,V_{tb}\Big[W^-_{\mu}\bar{b}\,\gamma^{\mu}
\left(f_1^L P_L +f_1^R P_R\right)t \nn\\ 
&&\qquad\qquad\qquad\qquad
\left.
-\frac{1}{m_W}\partial_{\nu}W^-_{\mu}\bar{b}\,\sigma^{\mu\nu}
  \left(f_2^L P_L +f_2^R P_R\right)t\right]\nn\\
&+&\frac{g}{\sqrt{2}}\,V_{tb}^*\Big[W^+_{\mu}\bar{t}\,\gamma^{\mu}
\left(\bar{f}_1^L P_L +\bar{f}_1^R P_R\right)b\nn\\
&&
\qquad\qquad\qquad\qquad\left.
-\frac{1}{m_W}\partial_{\nu}W^+_{\mu}\bar{t}\,\sigma^{\mu\nu}
  \left(\bar{f}_2^L P_L +\bar{f}_2^R P_R\right)b\right],
\eea
where $m_W$ is the mass of the $W$ boson,
$P_{L}=\frac{1}{2}(1-\gamma_5)$ and $P_{R}=\frac{1}{2}(1+\gamma_5)$ 
are the left- and right-handed chirality projectors, 
$\sigma^{\mu\nu}=\frac{i}{2}\left[\gamma^{\mu},
\gamma^{\nu}\right]$, $V_{tb}$ is the element of
the CKM matrix with the superscript * denoting 
complex conjugate, $f_{i}^{L}$, $f_{i}^{R}$, $\bar{f}_{i}^{L}$ and 
$\bar{f}_{i}^{R}$, $i=1,2$, are form factors which can be complex in general. 
Other dimension five terms that are possible in Lagrangian (\ref{wtblagr}) for 
off shell $W$ bosons vanish if the $W$'s decay into massless fermions, which
is a well satisfied approximation for fermions lighter than the $b$-quark.
The subroutines necessary for calculation of
the helicity amplitudes involving the right- and left-handed tensor
form factors of Lagrangian (\ref{wtblagr}) have been written and thoroughly
tested. 

The lowest order SM Lagrangian of the $Wtb$ interaction is reproduced 
by setting 
\bea
f_{1}^{L}=\bar{f}_{1}^{L}=1, \qquad 
f_{1}^{R}=f_{2}^{R}=f_{2}^{L}=
\bar{f}_{1}^{R}=\bar{f}_{2}^{R}=\bar{f}_{2}^{L}=0
\eea
in (\ref{wtblagr}).
If $CP$ is conserved then the following relationships hold
\bea
\left.\bar{f}_1^{R}\right.^*=f_1^R,
\quad \left.\bar{f}_1^{L}\right.^*=f_1^L \qquad 
{\rm and} \qquad
\left.\bar{f}_2^R\right.^*=f_2^L, \quad \left.\bar{f}_2^L\right.^*=f_2^R.
\eea
Thus, 4 independent form factors are left in Lagrangian (\ref{wtblagr}).
See \cite{frules} for the Feynman rules resulting from (\ref{wtblagr}).

Direct Tevatron limits,
obtained by investigating two form factors at a time
and assuming the other two at their SM values, are the
following \cite{D0} 
\bea
\left|f_{1}^R\right|^2 < 1.01, \qquad
\left|f_{2}^R\right|^2 < 0.23, \qquad
\left|f_{2}^L\right|^2 < 0.28.
\eea
The direct LHC limits are still weaker \cite{Aguilar}.
If $CP$ is conserved then the right-handed vector form factor and tensor 
form factors can be indirectly constrained from the CLEO data on
$b\rightarrow s\gamma$ branching fraction \cite{cleo} and from other rare $B$ 
decays \cite{fajfer}. However, there is 
still some room left within which the anomalous form factors,
in particular the tensor ones, can be varied.

The new version of {\tt carlomat} may be useful when looking for new 
physics effects in the processes of the top quark production both at 
hadron-hadron and $\epm$ collisions. In the context of the latter,
a feature of the program that may prove itself particularly useful 
is that full
information about helicities of the external particles can be easily
retrieved by switching off the MC summing over polarization states
of one or more particles.

\begin{figure}[htb]
\epsfig{file=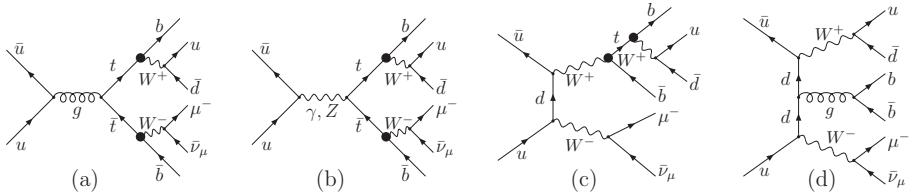,  width=120mm, height=25mm}
\caption{Examples of the lowest order Feynman diagrams of process 
(\ref{semilept}). Black blobs indicate the $Wtb$ coupling.}
\label{diags}
\end{figure}

Let us illustrate the usefulness of the automatic approach to the implementation
of the anomalous $Wtb$ coupling by looking closer at the top quark pair 
production in hadron-hadron collisions.
Main processes of $t\bar t$ production at hadron colliders are
\bea
q\bar{q} \rightarrow  t \bar{t},\qquad 
gg \rightarrow  t \bar{t}.
\eea
The quark-antiquark annihilation dominates at Tevatron 
while the gluon-gluon fusion dominates at LHC.
Taking into account decays $(t\to bW\to b f\bar{f}')$ results in processes
with 6 particles in the final state, as e.g.
\bea
\label{lept}
u\bar{u} &\ra& b \nu_{\mu}\mu^+\;\bar{b} \mu^-\bar{\nu}_{\mu},\\
\label{semilept}
u\bar{u} &\ra& b \nu_{\mu}\mu^+\;\bar{b} d\bar{u},\\
\label{hadr}
u\bar{u} &\ra& b u\bar{d}\;\bar{b} d\bar{u}
\eea
with 559, 718 and 6134 Feynman diagrams in the lowest order, respectively
(unitary gauge, $m_u=m_d=m_s=m_e=m_{\mu}=0$, no CKM mixing). Examples of
the Feynman diagrams of process (\ref{semilept}) are shown in Fig.~\ref{diags}.
Let us note that the $Wtb$ coupling that is indicated by a black blob enters
twice both in the $t\bar t$ production signal
diagrams depicted in Fig.~\ref{diags}a and \ref{diags}b
and the single top production diagram of Fig.~\ref{diags}c. Obviously it
is not present in the off resonance background diagrams
an example of which is shown in Fig.~\ref{diags}d. Needles to say that
the by hand implementation of the anomalous $Wtb$ coupling in the matrix
element of any of processes (\ref{lept})--(\ref{semilept}) would have
been rather tedious a task. With the current version of {\tt carlomat}
the cross section of these processes  can be computed
automatically with any choice of the anomalous form factors of (\ref{wtblagr}),
as it was done in \cite{afbtt}, where an influence of the anomalous 
$Wtb$ coupling on forward-backward asymmetry 
of top quark pair production at the Tevatron was investigated taking into
account decays of the top quarks to 6 fermion final states containing 
one charged lepton.

\section{Summary and Outlook}

New developments in {\tt carlomat} have been presented. They include
improvements of the generation of phase space integration routines,
implementation of scalar electrodynamics and the
anomalous $Wtb$ coupling including operators of dimension up to five and
many other improvements, as e.g., size reduction of the colour matrix,
that have not been discussed in the present lecture. 
Some minor bugs in the program have been corrected too.
A new version of the program will be released, hopefully soon.

Acknowledgements: This work was supported in part by the Research Executive 
Agency (REA) of the European Union under the Grant Agreement number 
PITN-GA-2010-264564 (LHCPhenoNet).

\end{document}